\documentclass[useAMS,usenatbib]{mn2e}

\usepackage{graphicx}
\usepackage{subfigure}
\usepackage{amsmath}
\usepackage{amssymb}

\title[Could AGN Outbursts Transform Cool Core Clusters?]{Could AGN Outbursts Transform Cool Core Clusters?} 
\author[F. Guo and S. P. Oh]{Fulai Guo$^{1,2}$\thanks{E-mail: fulai@ucolick.org} and S. Peng Oh$^{1}$\thanks{E-mail: peng@physics.ucsb.edu} \\
$^{1}$Department of Physics; University of California; Santa Barbara, CA 93106, USA\\
$^{2}$UCO/Lick Observatory, Department of Astronomy and Astrophysics, University of California, Santa Cruz, CA 95064, USA
}

\begin{document}
\bibliographystyle{mn2e}

\pagerange{000--000} \pubyear{0000}
\maketitle

\label{firstpage}

\begin{abstract}
The origin of the bimodality in cluster core entropy is still unknown. At the same time, recent work has shown that thermal conduction in clusters is likely a time-variable phenomenon. We consider if time-variable conduction and AGN outbursts could be responsible for the cool-core (CC), non cool-core (NCC) dichotomy. We show that strong AGN heating can bring a CC cluster to a NCC state, which can be stably maintained by conductive heating from the cluster outskirts. On the other hand, if conduction is shut off by the heat-flux driven buoyancy instability, then the cluster will cool to the CC state again, where it is stabilized by low-level AGN heating. Thus, the cluster cycles between CC and NCC states. In contrast with massive clusters, we predict the CC/NCC bimodality should vanish in groups, due to the lesser role of conductive heating there. We find tentative support from the distribution of central entropy in groups, though firm conclusions require a larger sample carefully controlled for selection effects. 
\end{abstract}

\begin{keywords}
galaxies: clusters: general -- cooling flows -- conduction -- galaxies: active -- instabilities -- X-rays: galaxies: clusters
\end{keywords}

\section{Introduction}
\label{section:intro}

There is a striking observational bimodality in the properties of galaxy cluster cores, which can be broadly separated into two types: cool-core (CC) and non cool-core (NCC) clusters. The former are defined to have temperature profiles which decline significantly toward the center. Measuring the relative abundance of the two types is somewhat hampered by selection effects (CC clusters have strongly peaked X-ray emission profiles which are more easily detected), but surveys indicate that roughly half of all clusters are in each category (e.g., \citealt{chen07}). The logarithmic slope of the entropy profile is bimodal, with CC/NCC clusters having steeper/shallower slopes respectively \citep{sanderson09}. The distribution of core entropy also appears to be bimodal in clusters, with population peaks at $K_{o} \sim 15 \, {\rm keV \, cm^{2}}$ and $K_{o} \sim 150 \, {\rm keV \, cm^{2}}$ and a distinct gap between $K_{o} \sim 30-50\, {\rm keV \, cm^{2}}$ \citep{cavagnolo09}. This in turns feeds into star formation and AGN properties: H$\alpha$ and radio emission from the central brightest cluster galaxy are much more pronounced when the cluster's core falls below an entropy threshold of $K_{o}< 30 \, {\rm keV \, cm^{-2}}$ \citep{cavagnolo08}, and a majority ($\sim 70\%$) of CC clusters host radio sources \citep{burns90}. 

Clearly, unravelling the origin of this dichotomy--for which there is no widely accepted explanation--could potentially yield great insight into cluster thermodynamics. Mergers have been considered a prime candidate for transforming CC to NCC systems, given the frequency of mergers in a hierarchical CDM cosmology, as well as the large amount of energy (as much as $\sim 10^{64}$ erg) in mergers, well in excess of that required. However, detailed simulations have not borne this expectation out. For instance, \citet{poole08} find that CC systems are remarkably robust and only disrupted in direct head-on or multiple collisions; even so, the resulting warm core state is only transient. To date, \citet{burns08} present the only set of simulations where NCC clusters are produced via mergers. For this to happen, nascent clusters must experience major mergers early on which destroyed embryonic CCs and prevented their reformation. Note that these simulations do not incorporate mechanisms (such as AGN feedback) to stop a cooling catastrophe; furthermore, the relatively low numerical resolution ($15.6 \, h^{-1} {\rm kpc}$) may preclude firm conclusions about core structure and evolution. As an alternative, \citet{mccarthy08} suggested that early pre-heating prior to cluster collapse could explain the lack of low entropy gas in NCC systems, which receive higher levels of preheating compared to CC systems. A possible concern in such scenarios is whether one can pre-heat the ICM to a high adiabat and yet retain sufficient low entropy gas in lower mass halos to obtain a realistic galaxy population (\citealt{bower08}; see also \citealt{oh_benson,evan_oh}). More importantly, many NCC clusters also have a short central cooling time ($\sim 1$ Gyr; \citealt{sanderson06}), and it is not clear why radiative cooling should not erase memory of the initial preheating episode. Both of these hypotheses focus on `nature', or initial conditions, in the form of an early major merger or preheating, in determining whether a cluster is CC or NCC. This hints at a fine-tuning problem: why are initial conditions such that roughly equal numbers of CC and NCC systems appear? In addition, there appears to be substantial differences in the metallicity profiles of NCC and CC systems, at least in groups (see discussion). Simulations show that metallicity profiles are remarkably stable to subsequent mergers \citep{poole08}. 

Recently, in \citet[hereafter GOR08]{guo08b}, we conducted a global Lagrangian stability analysis of clusters in which cooling is balanced by AGN heating and thermal conduction. This offered an alternative promising explanation, based on `nurture', or physical processes occurring in the ICM in its recent past. Our analysis showed that globally stable clusters could only exist in two forms: (1) cool cores stabilized by both AGN feedback and conduction, or (2) non-cool cores stabilized primarily by conduction\footnote{A recent study of a {\it Chandra} cluster sample has similarly found that while thermal conduction appears to be sufficient to stabilize NCC clusters, CC clusters appear to form a distinct population in which additional feedback heating is required \citep{sanderson09}}. Intermediate temperature profiles typically lead to globally unstable solutions, which would then quickly evolve to either CC or NCC states. In GOR08, we speculated that these two categories of clusters might even represent different stages of the same object. The importance of thermal conduction on global scales obviously depends on the large scale structure of the cluster magnetic fields. Recent calculations suggest that thermal conduction of heat into the cluster core can be self-limiting: in cases where the temperature decreases in the direction of gravity, a buoyancy instability (the heat flux driven buoyancy instability, hereafter HBI) sets in which re-orients a radial magnetic field to be largely transverse, shutting off conduction to the cluster center \citep{quataert08,parrish08a,parrish09,bogdanovic09}. In GOR08, we speculated (as subsequently did \citealt{bogdanovic09}) that powerful outbursts from a central AGN might counteract the HBI by disturbing the azimuthal nature of the magnetic field, thus enabling thermal conduction. In particular, the following scenario could arise: as conductivity falls, gas cooling and mass inflow will increase, triggering AGN activity. The rising buoyant bubbles may re-orient the magnetic field to be largely radial again, increasing thermal conduction and reducing mass inflow, shutting off the AGN until the HBI sets in once again. If AGN heating and/or thermal conduction during their `on' states are strong enough to heat the CC cluster to the NCC state, the cluster could then continuously cycle between cool-core (AGN heating dominated) and non cool-core (conduction dominated) states.  

The goal of this paper is to perform a very simple feasibility study for such a scenario, motivating future, more detailed work. While the transformation of NCC to CC clusters via radiative cooling can be easily accomplished (as seen, for instance, in Fig. 13 of \citet{parrish09}), the possible transformation of CC to NCC systems via AGN outbursts or time variable conduction has not been demonstrated. It has been conjectured before that strong AGN outbursts (the most extreme examples of which are Hydra A and MS 0435+7241) could permanently modify core entropy \citep{kaiser03,voit05}. But there have been no explicit calculations, and indeed suggestions that a CC to NCC transformation would be energetically prohibitive \citep{mccarthy08}. In this paper, we perform explicit calculations to investigate if strong AGN outbursts could transform a CC cluster to the NCC state (\S\ref{section:cycle}). Even if energetically subdominant (as they likely are for the most massive clusters), AGN could catalyze a dominant contribution from thermal conduction either due to: (i) the strong temperature dependence of thermal conductive flux, $F \propto T^{5/2}$; (ii) altering magnetic topology as discussed above. We study if such effects can indeed allow a CC to NCC transformation. It is important to note that the ability of rising bubbles or other bulk gas motions to globally restructure field geometry and hence thermal conductivity has not been demonstrated. However, there is important circumstantial support from numerical simulations of magnetic draping, which show that magnetic fields are amplified and more ordered in the wake of moving subhalos or bubbles \citep{ruszkowski07,asai07,dursi08,oneill09}. To isolate the relative contribution of AGN outbursts and conduction, we also consider models of galaxy groups. The strong temperature dependence of thermal conduction implies that conduction should be irrelevant in groups, regardless of magnetic field geometry. If our explanation for the bimodality in cluster cores is correct, then such bimodality should disappear in galaxy groups. 

We describe our methods in \S2, our results in \S3, and discuss implications in \S4. The cosmological parameters used throughout this paper are: $\Omega_{m}=0.3$, $\Omega_{\Lambda}=0.7$, $h=0.7$. We have rescaled observational results if the original paper used a different cosmology.

\section{Basic Assumptions and Setup}

We solve the time-dependent hydrodynamic equations using the ZEUS-3D hydrodynamic code \citep{stone92} in its one-dimensional mode; in particular, we have incorporated into ZEUS a background gravitational potential, radiative cooling, thermal conduction, convection, and AGN heating. 
We used the code similarly in \citet{guo08a}, albeit modified for the case of cosmic-ray heating, and gratefully acknowledge Mateusz Ruszkowski for supplying us with the modified version of ZEUS described in \citet[hereafter RB02]{ruszkowski02}, which was used as our base code. The model of AGN heating we adopt here is that of `effervescent heating' proposed by \citep{begelman01}, simulated in RB02, and also used in the global stability model of GOR08. We refer the reader to these papers for details of the models. Here we simply summarize several modifications and reiterate some important points to note. 

In all of the preceding papers, it was assumed that the AGN kinetic luminosity was directly related to the instantaneous mass accretion rate, $L = \epsilon \dot{\rm M} c^{2}$. This in itself was a simplification, given that AGN activity is likely to be intermittent, and not necessarily directly related to the mass inflow rate. It was justified on the grounds that AGN intermittency timescales are likely shorter than bubble rise times and gas cooling times, and hence that AGN feedback can be incorporated in a time-averaged sense. Here, we drop this assumption, and instead make the alternative (and perhaps more realistic) prescription of directly incorporating AGN outbursts and intermittency in the simulations. In our model, we assume that AGN is triggered when the central gas entropy drops below a critical value ($S_{5}< S_{\rm crit}$, where $S_{5}$ is the gas entropy at radius of $r=5$ kpc from the cluster center). We have also considered other AGN trigger criteria: for example, the AGN is triggered when the central mass inflow rate is larger than a critical value or the central gas cooling time is less than a critical value. We found that our results are fairly robust to the specific criterion we adopted. See Table \ref{table1} for the specific AGN trigger criterion adopted for each run. 

Once an AGN is triggered, we assume that the AGN heating lasts for a duration of order the bubble rise time, which is typically comparable to the sound crossing time $t_{\rm sc} \sim 10^{8} r_{100} c_{s,1000}^{-1}$yr for a radius $r \sim 100 r_{100}$ kpc and sound speed $c_{\rm s} \sim 1000 c_{s,1000} \, {\rm km \, s^{-1}}$ (e.g., see table 3 in \citet{birzan04}). Thus we assume that each AGN heating episode starts once the AGN trigger criterion is satisfied, and lasts for $t_{\rm agn}\sim 1\times 10^{8}$ yr. During the outburst, we assume a energy $E_{\rm agn} \sim 10^{60}-10^{61.5}$erg is liberated. Estimates of the work needed to inflate observed cavities in rich clusters yields $E_{\rm agn} \sim 10^{60}$erg \citep{birzan04,mcnamara05}, rising by 1-2 orders of magnitude in the most extreme events such as Hydro A \citep{nulsen05} and MS 0735+7421 \citep{mcnamara05}. By comparison, note that a $\sim 10^{9} \, {\rm M_{\odot}}$ black hole which doubles its mass over an accretion episode will liberate $\sim 10^{62} (\epsilon/0.05)$ erg, where $\epsilon$ is the efficiency in converting rest mass energy to kinetic energy. The AGN luminosity during each active AGN heating episode is then $L=E_{\rm agn}/t_{\rm agn}\sim 10^{44}-10^{46} \, {\rm erg \, s^{-1}}$. After a period $t_{\rm agn}$ with active AGN heating, we turn off the AGN heating, but continue to monitor the AGN trigger criterion. Once it is again satisfied, a new AGN heating episode starts. The AGN duty cycle (the period between the triggering of two successive AGN episodes) is set by the cooling time and is generally larger than $t_{\rm agn}$. During each outburst, the AGN heating rate is determined by the `effervescent heating' model described in RB02, except that here we adopt a stronger inner heating cutoff term $1-e^{-(r/r_{0})^{2}}$, instead of $1-e^{-r/r_{0}}$ adopted in RB02, to account for the finite size of the central radio source and to avoid overheating at the cluster centre. In the rest of this paper, the inner heating cutoff radius $r_{0}$ is taken to be $10$ kpc.

In our simulations, the gas entropy usually increases with radius, agreeing with observational trends. However, when the ICM is heated by a strong AGN outburst, negative entropy gradients may appear in some regions for a short time period. Thus, convection is also included in our calculation; the convective flux $F_{\rm conv}$ is given by the mixing length theory described in RB02. In cluster regions with negative entropy gradients, convection is turned on and transports thermal gas energy. Similar to RB02, we found that convection is not important for the parameters of the models presented in this paper: we ran our simulations for the same models but without convection and found similar results. In the low-density weakly-magnetized plasma (e.g., the ICM), anisotropic thermal conduction preferentially along the magnetic field lines modifies the usual Schwarzschild convective criterion through the magneto-thermal instability (MTI; \citealt{balbus00,parrish05,parrish07}). While we do not incorporate magnetic fields or MTI, recent calculations which do \citep{bogdanovic09,parrish09} similarly find it to be unimportant by two orders of magnitude.  

We do not conduct self-consistent MHD simulations of the HBI (as for instance in \citealt{parrish09,bogdanovic09}) and so instead employ a simplified toy model for the  conductivity. Given that the mechanism for overcoming the HBI is as yet unknown, we feel that such illustrative examples of the possible impact of time-variable conductivity is justified. We assume that radial conductivity is characterized by the Spitzer conductivity with some time-dependent suppression factor $f$. Initially, the conductivity is assumed to be negligible; we then assume that thermal conduction is efficient (see Table \ref{table1} for the value of $f$ adopted in each run) during each AGN outburst ($t_{\rm agn}\sim 1\times 10^{8}$ yr) and then is either turned off or decays as $\sim e^{-(t-t_{\rm off})/t_{\rm HBI}}$, where $t_{\rm off}$ is the end time of the preceding AGN heating episode. In practice, we have found that either assumption, as well as simulations where the onset of efficient conduction lags behind the AGN trigger, all yield very similar results with regard to the stability of the CC state. Here the HBI growth time $t_{\rm HBI}\sim 1.0\times 10^{8}$ yr for the typical cool core cluster A1795 and may be much larger for non-cool core clusters, since $t_{\rm HBI} \propto (d {\rm ln}T/dr)^{1/2}$ \citep{quataert08,parrish08}.

Our computational grid extends from $r_{\rm{min}}$ ($1$ kpc) to $r_{\rm{max}}$ ($200$ kpc for A1795 and $100$ kpc for NGC 4325). In order to resolve adequately the inner regions, we adopt a logarithmically spaced grid in which $(\Delta r)_{i+1}/(\Delta r)_{i}=(r_{\rm{max}}/r_{\rm{min}})^{1/N}$, where $N$ is the number of active zones. The standard resolution of our simulations presented in this paper is $N=400$; our code has been tested to be numerically convergent through simulations with different levels of resolutions. 

For initial conditions, we assume the ICM to be isothermal at $T=T_{\rm out}$, and solve for hydrostatic equilibrium. We assume that at the outer boundary $r_{\rm{max}}$, $n_{e}(r_{\rm max})=n_{\rm out}$, which is close to the value extrapolated from the observational density profile. For boundary conditions, we assume that the gas is in contact with a thermal bath of constant temperature and pressure at the outer radius, where the cooling time exceeds the Hubble time. Thus, we ensure that temperature and density of the thermal gas at the outer radius are constant. We extrapolate all hydrodynamic variables from the active zones to the ghost zones by allowing them to vary as a linear function of radius at both the inner and outer boundaries. The intracluster gas is allowed to flow in and out of active zones at both the inner and outer boundaries.

\section{Results}

\subsection{Stability in the CC state}
\label{section:stability}

\begin{figure}
\includegraphics[width=0.45\textwidth]{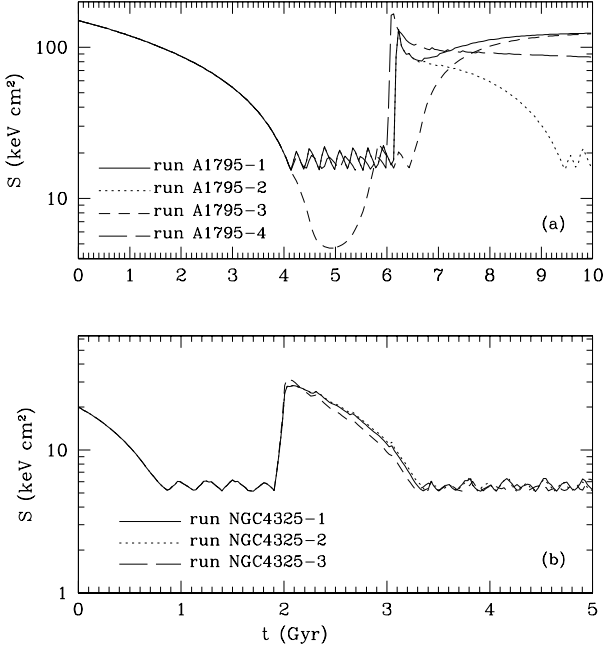}
\caption{Time evolution of entropy at $r=5$ kpc in various models for the cluster Abell 1795 ({\it top}) and for the group NGC 4325 ({\it bottom}). A strong AGN outburst and/or a sharp increase in conductivity is able to accomplish a transition from a CC to NCC state.}
 \label{plot1}
 \end{figure}

Let us begin by considering how clusters can be stabilized in the CC state by a combination of AGN and conductive heating. We first run simulations for a typical massive cluster Abell 1795. The parameters for this cluster are $M_{0} = 6.6 \times 10^{14}$ M$_{\sun}$, $r_{\rm{s}}=460$ kpc, $r_{\rm{c}}=r_{\rm{s}}/20$ (\citealt{zakamska03}; also see GOR08 for details), $T_{\rm out}=6.8$ keV, and $n_{\rm out}=0.003$ cm$^{-3}$. Specific model parameters for each run are listed in Table \ref{table1}.

Run A1795-1 is a representative simulation. The cluster is initially in hydrostatic equilibrium with spatially constant temperature $T=T_{\rm out}=6.8$ keV, and then evolves via radiative cooling without AGN heating or thermal conduction. The solid line in the {top} panel of Figure \ref{plot1} shows the time evolution of gas entropy at $r=5$ kpc ($S_{5}$), which decreases gradually in the first $4$ Gyr. When the central gas entropy drops below $15$ keV cm$^{2}$ (at $t\sim 4$ Gyr), an AGN outburst with $E_{\rm agn}=2\times 10^{60}$ erg is triggered and lasts for $t_{\rm agn}= 1\times 10^{8}$ yr. During the period of active AGN heating, thermal conduction with $f=0.4$ is also turned on. As clearly shown in Fig. \ref{plot1}, the cooling catastrophe is quickly averted, and the gas entropy increases. After $0.1$ Gyr of active heating, both AGN heating and thermal conduction are turned off and the cluster cools again until the next heating episode is triggered. As seen for $t\sim 4-6$ Gyr in Fig. \ref{plot1} (top panel), the ICM entropy executes minor oscillations and the cluster stays in the cool core state, where radial profiles of gas temperature and density fit observational data \citep{ettori02} very well. The radial profiles of entropy oscillating in the CC state are shown in Figure \ref{plot2} (top), where entropy profiles are plotted every $0.1$ Gyr since $t= 4$ Gyr. During the CC state, the lines are clearly concentrated in the lower branch with central gas entropy $\sim 10-20$ keV cm$^{2}$.

In run A1795-1, the AGN duty cycle is $\sim 0.3$ Gyr. During each heating episode, the volume-integrated conductive heating energy is around $2E_{\rm agn}$, i.e., the conductive heating is comparable to AGN heating. We have also done similar calculations for the cluster A2199 ($T_{\rm out}=4.6$ keV) and the group NGC 4325 ($T_{\rm out}=1$ keV), and found that conductive heating is an order of magnitude smaller than AGN heating in the former and becomes negligible in the latter. The results for NGC 4325 are also presented in this paper for comparison. The parameters for this group are $M_{0} = 1.1 \times 10^{13}$ M$_{\sun}$, $r_{\rm{s}}=78.3$ kpc, $r_{\rm{c}}=0$ kpc, $T_{\rm out}=1$ keV, and $n_{\rm out}=7.27 \times 10^{-4}$ cm$^{-3}$ \citep{2007ApJ...669..158G}. As can be seen (lower panel Fig \ref{plot1}, $t \sim 0.8-2$ Gyr), the group is similarly stabilized in the CC state with small entropy oscillations. 

We have done our calculations with different levels of AGN heating and thermal conduction, and found that our results are quite robust. Higher levels of AGN heating usually correspond to larger amplitude entropy oscillations in the CC state. We also consider a model (run A1795-3) where AGN feedback only triggers conduction without heating the ICM (i.e., $E_{\rm agn}=0$ erg), and find that while the cluster cools significantly to low central entropy, it does not end up in a cooling catastrophe, and instead eventually ends up as a CC cluster as well (see the short-dashed line in Fig. \ref{plot1}a when $t\lesssim6.4$ Gyr). Note that conduction is regulated by AGN feedback in this run before $t\sim6.4$ Gyr, after which it is fixed to be $f=0.4$ without regulation (see further discussion in \S\ref{section:cycle}).

Although present in above runs (to account for the shut-off of conductivity by the HBI), the regulation of conductivity is not required for the stability of the CC state. In run A1795-4, we considered a model where conductivity is triggered and then fixed to be $f=0.2$ without regulation since $t\sim 4$ Gyr; the fixed lower value is meant to simulate a situation where the field line geometry reaches a steady state balance between the HBI and some other mechanism (e.g, stirring by galaxies). The cluster evolution is similar to other runs. Note that for this lower value of $f$, the cluster {\it would} reach a cooling catatrophe without AGN feedback heating. In addition, for a fixed value of $f$, lower temperature clusters, and galaxy groups usually reach cooling catastrophes if only thermal conduction operates. 

Note that a minimum amount of thermal conduction is usually required in our calculations, since AGN ``effervescent heating" tends to be excessively centrally concentrated: the very central regions of the cluster are overheated, while outer regions develop a cooling catastrophe. Furthermore, thermal conduction is the dominant energy source in  massive high-temperature clusters; without it, a much larger $E_{\rm agn}$ (inconsistent with observations) is required to maintain the ICM in the CC state. On the other hand, in low temperature groups such as NGC 4325, AGN heating alone suffices, as seen in the lower panel of Fig. \ref{plot1}. 

\subsection{The cycle between CC and NCC states}
\label{section:cycle}

 \begin{figure}
\includegraphics[width=0.45\textwidth]{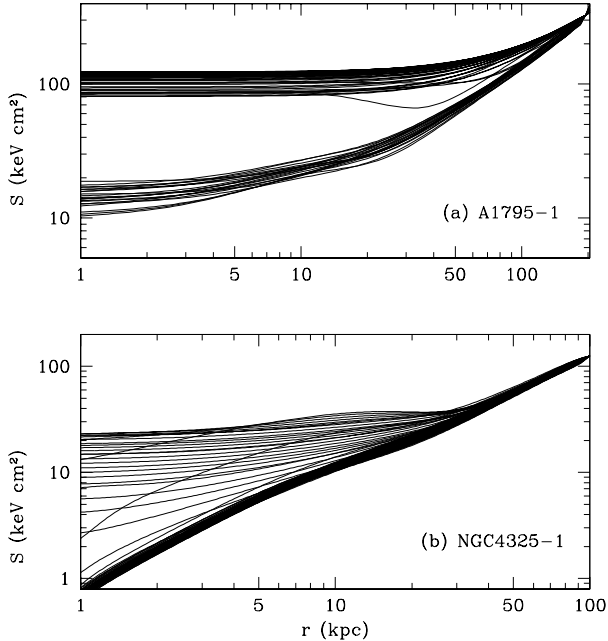}
\caption{Time sequence of entropy in a typical model for the cluster Abell 1795 (run A1795-1 from $t=4$ Gyr to $10$ Gyr; {top} panel) and for the group NGC 4325 (run NGC4325-1 from $t=1$ Gyr to $5$ Gyr; {bottom} panel). Each line in the {top} panel is plotted every $0.1$ Gyr and that in the {bottom} panel is plotted every $0.05$ Gyr. The cluster A1795 oscillates in the CC state (lower branch) due to sporadic AGN feedback heating before $t\sim 6.1$ Gyr, when the cluster is quickly heated by a strong AGN outburst to the NCC state. The lines then concentrate in the NCC state (upper branch). The evolution of Abell 1795 is clearly bimodal, while bimodality in the evolution of NGC 4325 is not evident.}
 \label{plot2}
 \end{figure}

Strong AGN outbursts, e.g., Hercules A \citep{nulsen05} and MS0735.6+7421 (\citealt{mcnamara05}; \citealt{mcnamara09}), with $E_{\rm agn}$ up to $\sim 10^{62}$ ergs have been found in X-ray observations. In this subsection, we consider the impact of such strong AGN outbursts on the evolution of cool core clusters. We did our calculations for both the cluster A1795 and the group NGC 4325, aiming to investigate if such strong AGN outbursts can heat the CC system into a NCC state and what the role of thermal conduction is in this context. 

In run A1795-1, the first AGN outburst after $t=6$ Gyr is modified to be much stronger ($E_{\rm agn}=3\times 10^{61}$ erg). Figure \ref{plot1}(a) clearly shows that the central gas entropy is quickly boosted from $\sim 15$ keV cm$^{2}$ to $\sim 130$ keV cm$^{2}$ at $t\sim6.1$ Gyr. The cluster indeed reaches a NCC state; the radial entropy profile is shown in the upper branch of Figure \ref{plot2}(a). During the strong AGN outburst, AGN heating is much stronger (by one order of magnitude) than conductive heating. However, given the spatial dependence of AGN heating, conduction is still important in transporting energy within the cluster to offset cooling in certain regions. Since the cluster's temperature profile in the NCC state is nearly isothermal, the HBI timescale, which scales as $(d {\rm ln}T/dr)^{1/2}$ \citep{quataert08}, is very long. We therefore assume that for run A1795-1, thermal conduction in the NCC state does not decay. Figure \ref{plot2}(a) clearly shows that the ICM adjusts itself to the NCC state where cooling is balanced by thermal conduction alone. Global stability of such conduction only models is possible in the hottest clusters for a high level of conductivity (GOR08). 

Run A1795-3 is a model where AGN feedback only triggers conduction without heating the ICM directly (i.e., $E_{\rm agn}=0$ erg). When $t\lesssim6.4$ Gyr, the cluster reaches and is then maintained in the CC state by AGN-regulated conduction (see \S\ref{section:stability}). After $t\sim6.4$ Gyr, we assume that AGN is continuously active so that conduction with $f=0.4$ does not decay with time. The short-dashed line in Figure (\ref{plot1})a clearly shows that the massive cluster A1795 is also heated to the NCC state, though this process takes much longer than that in run A1795-1. This shows that in the hottest clusters, if AGN can trigger a high level of conductivity, their heat input is unimportant. This is not true if the triggered level of conductivity is lower. In run A1795-4, we assume that conductivity is triggered and then fixed to be $f=0.2$ since $t\sim4$ Gyr. The cluster evolution is similar to other runs, with the strong AGN outburst bringing the cluster to a NCC state. However, if the AGN contributes negligible heat input, as in run A1795-3, the lower level of conduction in this run would be unable to prevent a cooling catastrophe.  

Since thermal conduction will eventually decay due to the HBI, the cluster is expected to cool from the NCC state to the CC state. The HBI linear growth time for the CC cluster A1795 is $t_{\rm HBI}\sim 1.0\times 10^{8}$ yr; however, in the NCC state, the temperature profile is very flat and thus the HBI growth time, which scales as $(d {\rm ln}T/dr)^{1/2}$ \citep{quataert08}, becomes much longer. Furthermore, simulations show that the cluster takes several instability growth times for the magnetic field lines to be appreciably re-oriented. Thus, in run A1795-2, we assume that conduction decays on a timescale of $t_{\rm HBI}=1$ Gyr after the strong AGN outburst is turned off. The dotted line in Figure (\ref{plot1})a demonstrates that the cluster cools gradually until triggering new AGN activity and then stays in the CC state. Thus, the cluster cycles between the CC state and NCC state due to strong AGN outbursts and the HBI.

For the group NGC 4325, we modify the AGN outburst at $t\sim 1.9$ Gyr to be much stronger ($E_{\rm agn}=2\times 10^{59}$ erg) and found that the group is heated to the NCC state. Figure \ref{plot1}b clearly shows that the group cycles between the CC state and NCC state for all three models considered in this paper (see Table \ref{table1} for parameters in each model), with the key difference that unlike the cluster case, the group does not stay in the NCC state for long. We have also done our calculations for the cluster Abell 2199 ($T_{\rm out}=4.6$ keV) and found that a strong AGN outburst with $E_{\rm agn}=5\times 10^{60}$ erg is able to heat the CC cluster to its NCC state. Thus , our calculations suggest the trend that more massive systems require stronger AGN outbursts to reach the NCC state.

From Figure (\ref{plot1}), a scenario for the evolution of galaxy clusters and groups may arise naturally: a cool core group or cluster is maintained by normal AGN feedback heating, and may be heated by a strong AGN outburst to the NCC state, from which the system may gradually cool to the CC state again due to the decay of conductivity by the HBI. The strong AGN outburst may be triggered by a sudden increase in close gas supply, or mergers\footnote{Indeed, little is know about the mechanical variability of AGN; possible reasonable assumptions are that the luminosity has a log-normal distribution with a 'flicker-noise' power spectrum \citep{nipoti05}. If our hypothesis is correct, then demography, in particular the relative fraction of CC and NCC systems, may hold the key to understanding the frequency of outbursts as a function of energy.}. This scenario naturally explains current X-ray observations of both CC and NCC groups and clusters. The typical timescale of the HBI is around $0.1$ Gyr for the cool core state of A1795 and varies as $t_{\rm HBI} \propto (d {\rm ln}T/dr)^{1/2}$ \citep{quataert08} during the cluster evolution. Clusters with flatter temperature profiles usually have larger $t_{\rm HBI}$, and thus clusters with quite flat temperature profiles may dominate in the population of NCC clusters. This agrees with current X-ray cluster observations (see Fig. 6 of \citealt{sanderson06}).

Calculations by \citet{mccarthy08} suggest that heating a pure cooling flow cluster to a NCC cluster requires extremely large amounts of energy ($\sim 10^{63}$ erg). In our calculations, AGN feedback is triggered long before strong cooling flows form, and thus the needed AGN energy is much less. We have performed simulations where AGN outbursts are triggered when a strong mass inflow forms (``the cooling flow state"), and found that much larger (several times) AGN heating is required to heat the cluster to its NCC state. The absence of strong cooling flows in clusters suggests that the cluster stabilizes at a CC state far before strong cooling flows form. This permits a NCC state to be attained by heating from a CC state, as we have seen in our simulations.

\subsection{The bimodality of CC and NCC states}
\label{section:bimodality}

 \begin{figure}
\includegraphics[width=0.45\textwidth]{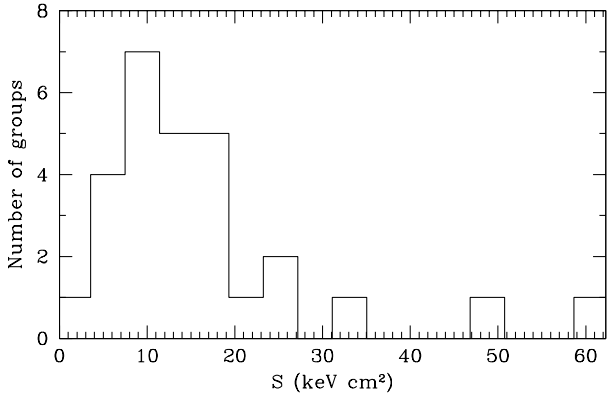}
\caption{Histogram of gas entropy at $0.01r_{500}$ for a sample of 28 nearby galaxy groups from the Two-Dimensional {\it XMM-Newton} Group Survery \citep{johnson09}. If the selection effect is not important, the histogram suggests that the group distribution is unimodal with only one peak for CC groups. }
 \label{plot3}
 \end{figure}

Both CC and NCC groups and clusters have been observed in nature. Recently, \citet{sanderson09} and \citet{cavagnolo09} demonstrate that the CC/NCC bimodality does exist in clusters: Figure 12 in the former shows the bimodality in the distribution of the logarithmic slopes of radial entropy profiles, while Figure 6 and 7 in the latter demonstrate the bimodality in the distribution of central gas entropy.

In Figure \ref{plot2}(a), radial entropy profiles in run A1795-1 are plotted every $0.1$ Gyr since $t= 4$ Gyr, when the ICM reaches the CC state (see Fig. \ref{plot1}a). As clearly seen, the ICM goes through minor oscillations in the CC state before $t\sim 6.1$ Gyr (lines concentrate in the lower branch). When the strong AGN outburst is triggered at $t\sim 6.1$ Gyr, the cluster is then quickly heated to the NCC state, and stays there due to conductive heating in the NCC state: the lines concentrate there (upper branch). We clearly see the bimodality in the cluster evolution. Thus our simulations suggest that strong AGN outbursts may heat the CC cluster to its NCC state and the cluster CC/NCC bimodality naturally appears in this scenario.

In run A1795-1, conduction is regulated by AGN feedback and the HBI. In contrast, we have also considered a model where conduction is fixed (uncorrelated with AGN activity; run A1795-4). As shown in Figure \ref{plot1}(a), we found that the cluster is also heated by the strong AGN outburst at $t\sim 6$ Gyr to the NCC state, where the cluster then stays due to conductive heating during our whole simulation. It seems that the regulation of thermal conduction is not necessary for the CC/NCC bimodality. However, note that in fixed-conductivity models, $f$ must has a value within a small range: if $f$ is too large, the cluster will evolve to the NCC state without staying in the CC state (e.g. run A1795-3 after $t\sim6.4$ Gyr); if $f$ is too small, the cluster may not stay in the NCC state, but instead develop a cooling catastrophe. In other words, if $f$ is too large, we will not see a significant population of massive CC clusters; if $f$ is too small, we will not see a significant population of NCC clusters. On the other hand, time-varying conductivity regulated by the AGN could naturally circumvent this 'fine-tuning' problem: the cluster can stay in the CC state due to the alternation of radiative cooling and intermittent heating by AGN feedback and conduction; the cluster can also stay in the NCC state, where efficient conduction triggered by strong AGN outbursts offsets radiative cooling. Furthermore, since conductivity increases strongly with temperature, higher temperature clusters may stay in the NCC state for a much longer time, which is consistent with the observational finding that the fraction of NCC systems in clusters increases with the cluster mass \citep{chen07}.

Figure \ref{plot2}(b) shows the same plot, but for the $1$ keV group NGC 4325 (run NGC4325-1). The lines are shown every $0.05$ Gyr from $t= 1$ Gyr, when the ICM is in the CC state (see Fig. \ref{plot1}b) to $t= 5$ Gyr. At $t\sim1.9$ Gyr, the group is heated by a strong AGN outburst to the NCC state. However, since the conductive heating is inefficient in low-temperature systems ($\kappa \propto T^{5/2}$), the ICM can not be maintained in the NCC state due to radiative cooling and cools to the CC state again (also see Fig. \ref{plot1}). In run NGC4325-3, conductivity triggered by the strong AGN outburst is modified to be as large as the Spitzer value (see Table 1) and we still found that the group evolution is similar to that in run NGC4325-1 (see Fig. \ref{plot1}b). We tried to build an equilibrium model with conduction alone for the NCC group NGC 4325 and found that conduction with 10 times Spitzer value is required to balance radiative cooling. Obviously, in Figure (\ref{plot2})b, the CC/NCC bimodality is not evident in the group evolution; instead, the group distribution is unimodal: lines are only concentrated in the CC state, and the group distribution in NCC states is continuous rather than peaked.
If strong AGN outbursts are not common, NCC groups may be rare, since they can not be maintained in NCC states by conductive heating. 

To test our model, we turn to group observations in literature. In \citet{rasmussen07}, 14 of their 15 groups observed by {\it Chandra} are in the CC state, which may be due to a selection effect, since all of their groups are reasonably X-ray bright. Here we adopt a sample of $28$ nearby groups studied by \citet{johnson09}, which is the largest sample to date with high-quality {\it XMM-Newton} data. There exist large Chandra samples, but these only select bright groups. We plot the distribution of the central gas entropy at $0.01r_{500}$ in Figure (\ref{plot3}), where $r_{500}$ is the radius within which the mean density of the group is $500$ times the critical density. Figure \ref{plot3} clearly shows that the central entropy distribution is unimodal and that groups with high values of central entropy ($S\gtrsim 20$ keV cm$^{2}$) are rare. Thus current group observations seem to agree with our results. However, note that NCC groups are very faint, and many of them may have not been observed yet. If more NCC groups are observed in future, we predict that their distribution is continuous, instead of peaked around one specific NCC state. A large group sample carefully controlled for selection effects would be required to do a more reliable test of our model.

\begin{table*}
 \centering
 \begin{minipage}{160mm}
  \renewcommand{\thefootnote}{\thempfootnote} 
  \caption{List of Simulations.}
    \vspace{0.1in}
  \begin{tabular}{@{}lcccccccc}
  \hline
            & {$S_{\rm crit}$\footnote{Each AGN heating episode is triggered when gas entropy ($S\equiv k_{\rm B}T/ n_{\rm{e}}^{2/3}$) at $r=5$ kpc drops below $S_{\rm crit}$.}}& {$E_{\rm agn}$\footnote{The mechanical energy released during a weak ($E_{\rm agn}$) or strong ($E_{\rm agn,s}$) AGN outburst. We assume that each AGN outburst heats the ICM for a duration of $t_{\rm agn}=1.0\times 10^{8}$ yrs.} }& {$ f$\footnote{The conduction suppression factor relative to the Spitzer value when AGN heating ($f$) or a strong AGN outburst ($f_{s}$) is active.}} & {$t_{\rm HBI}$\footnote{Conduction is on during each active AGN heating episode. When AGN is turned off, conductivity then decays exponentially in a timescale of $t_{\rm HBI}$ (after weak AGN outbursts) or $t_{\rm HBI,s}$ (after strong AGN outbursts). $t_{\rm HBI}=\infty$ indicates non-decaying conduction, while $t_{\rm HBI}=0$ indicates that conduction is turned off once AGN heating is shut off.}}&  {$E_{\rm agn,s}^{\mbox{\it b}}$} &{$f_{\rm s}^{\mbox{\it c}}$} & {$t_{\rm HBI,s}^{\mbox{\it d}}$}\\
      Run& keV cm$^{2}$  & $(10^{60}$ erg)&&(Gyr)&$(10^{60}$ erg)&&(Gyr)\\
 \hline
A1795-1 &15& 2.0 & 0.4&0 & 30 &0.4&$\infty$     \\
A1795-2 & 15& 2.0 & 0.4 &0&  30 &0.4& 1      \\
A1795-3 & 15& 0.0 & 0.4 & 0& 0 &0.4&$\infty$      \\
A1795-4 & 15& 2.0 & 0.2 &$\infty$&  30 &0.2&$\infty$      \\
NGC4325-1 &5&  0.03 & 0.4 &0&  0.2 &0.4& $\infty$     \\
NGC4325-2& 5& 0.03 & 0.4 & 0& 0.2 &0.4& 1     \\
NGC4325-3 &5&  0.03 & 0.4 &0&0.2 &1.0& $\infty$     \\
 \hline
\label{table1}
\end{tabular}
\end{minipage}
\end{table*}

\section{Discussion}

Let us briefly summarize our findings. From a suite of 1D hydrodynamic simulations, we find that clusters can cycle between CC and NCC states, driven by time-variable conduction and/or AGN outbursts. A strong AGN outburst combined with conduction could heat a CC group or cluster to the NCC state. During this transition, AGN usually provides most of the heating energy, while conduction is important in transporting energy within the cluster to offset cooling in certain regions, given the spatial dependence of AGN heating. The relative importance of conduction increases with cluster temperature, due to the strong temperature dependence of conductive flux, $F \propto T^{5/2}$. High temperature clusters, provided that conduction in the `on' state is relatively unsuppressed $f \sim 0.4$ for an extended time, can even reach the NCC state with no energy input from the AGN. In this case, the AGN may simply serve as a `switch' to regulate conductivity, perhaps by straightening field lines via the production of rising bubbles. Lower temperature clusters (or high temperature clusters if the maximum value of conduction is still relatively weak) require a combination of AGN and conductive heating to attain the NCC state. In both cases, if conduction continues to operate, the cluster can remain stably in the NCC state. On the other hand, if conduction decays via the HBI, then the cluster will cool and revert back to the CC state, where it remains stably with normal AGN feedback (\S\ref{section:stability}) until the next strong outburst and/or strong increase in conductivity continues the cycle. At the low temperature end, groups cannot be stabilized by any means in the NCC state, and rapidly cool to the CC state until the next outburst. The duty cycle, or timescale to cycle between CC and NCC states, shortens with declining temperature. 

If this hypothesis for the origin of CC/NCC cluster cores is correct, a number of interesting conclusions follow. Since the stabilizing effects of conduction decline with temperature, the NCC/CC bimodality should be a function of temperature, being the most sharply defined for high temperature clusters, and vanishing in galaxy groups. We show there may be tentative observational evidence for lack of bimodality in core entropy in the group sample of \citet{johnson09}, although selection effects have to be carefully quantified before one can draw firm conclusions. The relative abundance of CC/NCC clusters may gives insights on the duty cycle on which conduction (and/or AGN outbursts) vary. For instance, the roughly equal fraction of CC/NCC clusters suggests that the AGN duty cycle between strong outbursts in CC systems is of order the HBI/cooling timescale in the corresponding NCC systems, though a more detailed study attempting to match the fraction of time a cluster spends at a given central entropy, to the distribution of entropies in the cluster population as a whole, would be interesting. Once again, groups can perform a critical test, since they will not spend much time in a high-entropy state. Furthermore, since the duty-cycle between strong AGN outbursts is shorter in groups, turbulence or convective effects due to AGN activity which leave an imprint on the metallicity or entropy profile might have a more pronounced effect there. One possible example is the turbulent diffusion of metals \citep{rebusco05,sharma09}. In addition to heating the CC cluster to the NCC state, strong AGN outbursts could potentially remove the centrally-peaked metalicity distribution observed in CC systems, resulting in a relatively flat metallicity profile in the NCC state. While this distinction was seen in \citet{de-grandi01}, more recent studies by \citet{baldi07}, \citet{leccardi08}, and \citet{sanderson09} found that outside the very innermost regions, metallicity profiles were consistent with a single power-law at all radii for both CC and NCC clusters. In contrast, the metallicity profiles in NCC groups are much flatter than those in CC groups \citep{johnson09b}. This is consistent with the model presented in this paper. In particular, NCC clusters can be stabilized by thermal conduction for sufficiently long periods for the metallicity gradient to be re-established, while NCC groups have a much more recent origin due to the short duty cycle of strong AGN outbursts, and thus retain evidence of AGN `stirring' in the metallicity profile.\footnote{We thank Trevor Ponman for pointing this out.}

Our 1D calculations are frankly exploratory and simplified in nature. Our hope is to show that AGN outbursts and time-variable conductivity are a plausible means of regulating the bimodality between NCC and CC systems, motivating future, more detailed work. Of course, the greatest gap in our understanding is the actual means by which the HBI can be counteracted to allow thermal conduction to operate. Even if the rising bubbles do not cause a radial reorientation of field lines, as we have suggested, {\it some} mechanism (perhaps stirring of the gas by galaxies or subhalos) must be counteracting the onset of the HBI in clusters; otherwise, conduction is no longer a viable heating mechanism. This alone would require significant revision of theoretical models, since no known heating mechanisms (such as AGN heating or dynamical friction) acting alone without conduction is sufficient to offset a cooling catastrophe in massive clusters (e.g, see \citealt{conroy08}): such mechanisms tend to be too centrally concentrated toward the core, and only marginally sufficient energetically. Furthermore, it would then seem a remarkable coincidence that if one simply uses observed temperature profiles in clusters to construct the Spitzer conductive flux from the cluster outskirts, it is very nearly equal to that required to balance the radiative cooling rate as indicated by the observed X-ray surface brightness profile, for some reasonable fraction $f\sim 0.3$ of the Spitzer value (e.g., Fig. 17 of \citealt{peterson_fabian}). There is no reason in principle why such close agreement should exist, and seems a tantalizing hint that nature somehow `knows' about Spitzer conductivity. Much work remains to be done before we understand if there are large secular variations to the apparent thermal equilibrium in clusters. 

\section*{Acknowledgments}
We thank the referee, Chris Reynolds, for a helpful report. FG thanks Trevor Ponman and Ewan O'Sullivan for helpful discussions and William Mathews for a careful reading of the manuscript. SPO thanks MPA for hospitality. We acknowledge support by NASA grant NNG06GH95G. 

\bibliography{ms} 

\begin{thebibliography}{}

\bibitem[\protect\citeauthoryear{{Asai}, {Fukuda} \& {Matsumoto}}{{Asai}
  et~al.}{2007}]{asai07}
{Asai} N.,  {Fukuda} N.,    {Matsumoto} R.,  2007, \apj, 663, 816

\bibitem[\protect\citeauthoryear{{Balbus}}{{Balbus}}{2000}]{balbus00}
{Balbus} S.~A.,  2000, \apj, 534, 420

\bibitem[\protect\citeauthoryear{{Baldi}, {Ettori}, {Mazzotta}, {Tozzi} \&
  {Borgani}}{{Baldi} et~al.}{2007}]{baldi07}
{Baldi} A.,  {Ettori} S.,  {Mazzotta} P.,  {Tozzi} P.,    {Borgani} S.,  2007,
  \apj, 666, 835

\bibitem[\protect\citeauthoryear{{Begelman}}{{Begelman}}{2001}]{begelman01}
{Begelman} M.~C.,  2001, in {Hibbard} J.~E.,  {Rupen} M.,   {van Gorkom} J.~H.,
   eds, ASP Conf. Ser. Vol. 240, Gas and Galaxy Evolution. {Astron. Soc. Pac.,
  San Francisco}.
p.~363

\bibitem[\protect\citeauthoryear{{B{\^\i}rzan}, {Rafferty}, {McNamara}, {Wise}
  \& {Nulsen}}{{B{\^\i}rzan} et~al.}{2004}]{birzan04}
{B{\^\i}rzan} L.,  {Rafferty} D.~A.,  {McNamara} B.~R.,  {Wise} M.~W.,
  {Nulsen} P.~E.~J.,  2004, \apj, 607, 800

\bibitem[\protect\citeauthoryear{{Bogdanovic}, {Reynolds}, {Balbus} \&
  {Parrish}}{{Bogdanovic} et~al.}{2009}]{bogdanovic09}
{Bogdanovic} T.,  {Reynolds} C.~S.,  {Balbus} S.~A.,    {Parrish} I.~J.,  2009,
  preprint (arXiv: 0905.4508)

\bibitem[\protect\citeauthoryear{{Bower}, {McCarthy} \& {Benson}}{{Bower}
  et~al.}{2008}]{bower08}
{Bower} R.~G.,  {McCarthy} I.~G.,    {Benson} A.~J.,  2008, \mnras, 390, 1399

\bibitem[\protect\citeauthoryear{{Burns}}{{Burns}}{1990}]{burns90}
{Burns} J.~O.,  1990, \aj, 99, 14

\bibitem[\protect\citeauthoryear{{Burns}, {Hallman}, {Gantner}, {Motl} \&
  {Norman}}{{Burns} et~al.}{2008}]{burns08}
{Burns} J.~O.,  {Hallman} E.~J.,  {Gantner} B.,  {Motl} P.~M.,    {Norman}
  M.~L.,  2008, \apj, 675, 1125

\bibitem[\protect\citeauthoryear{{Cavagnolo}, {Donahue}, {Voit} \&
  {Sun}}{{Cavagnolo} et~al.}{2008}]{cavagnolo08}
{Cavagnolo} K.~W.,  {Donahue} M.,  {Voit} G.~M.,    {Sun} M.,  2008, \apjl,
  683, L107

\bibitem[\protect\citeauthoryear{{Cavagnolo}, {Donahue}, {Voit} \&
  {Sun}}{{Cavagnolo} et~al.}{2009}]{cavagnolo09}
{Cavagnolo} K.~W.,  {Donahue} M.,  {Voit} G.~M.,    {Sun} M.,  2009, \apjs,
  182, 12

\bibitem[\protect\citeauthoryear{{Chen}, {Reiprich}, {B{\"o}hringer}, {Ikebe}
  \& {Zhang}}{{Chen} et~al.}{2007}]{chen07}
{Chen} Y.,  {Reiprich} T.~H.,  {B{\"o}hringer} H.,  {Ikebe} Y.,    {Zhang}
  Y.-Y.,  2007, \aap, 466, 805

\bibitem[\protect\citeauthoryear{{Conroy} \& {Ostriker}}{{Conroy} \&
  {Ostriker}}{2008}]{conroy08}
{Conroy} C.,  {Ostriker} J.~P.,  2008, \apj, 681, 151

\bibitem[\protect\citeauthoryear{{De Grandi} \& {Molendi}}{{De Grandi} \&
  {Molendi}}{2001}]{de-grandi01}
{De Grandi} S.,  {Molendi} S.,  2001, \apj, 551, 153

\bibitem[\protect\citeauthoryear{{Dursi} \& {Pfrommer}}{{Dursi} \&
  {Pfrommer}}{2008}]{dursi08}
{Dursi} L.~J.,  {Pfrommer} C.,  2008, \apj, 677, 993

\bibitem[\protect\citeauthoryear{{Ettori}, {Fabian}, {Allen} \&
  {Johnstone}}{{Ettori} et~al.}{2002}]{ettori02}
{Ettori} S.,  {Fabian} A.~C.,  {Allen} S.~W.,    {Johnstone} R.~M.,  2002,
  \mnras, 331, 635

\bibitem[\protect\citeauthoryear{{Gastaldello}, {Buote}, {Humphrey},
  {Zappacosta}, {Bullock}, {Brighenti} \& {Mathews}}{{Gastaldello}
  et~al.}{2007}]{2007ApJ...669..158G}
{Gastaldello} F.,  {Buote} D.~A.,  {Humphrey} P.~J.,  {Zappacosta} L.,
  {Bullock} J.~S.,  {Brighenti} F.,    {Mathews} W.~G.,  2007, \apj, 669, 158

\bibitem[\protect\citeauthoryear{{Guo} \& {Oh}}{{Guo} \& {Oh}}{2008}]{guo08a}
{Guo} F.,  {Oh} S.~P.,  2008, \mnras, 384, 251

\bibitem[\protect\citeauthoryear{{Guo}, {Oh} \& {Ruszkowski}}{{Guo}
  et~al.}{2008}]{guo08b}
{Guo} F.,  {Oh} S.~P.,    {Ruszkowski} M.,  2008, \apj, 688, 859

\bibitem[\protect\citeauthoryear{{Johnson}, {Ponman} \& {Finoguenov}}{{Johnson}
  et~al.}{2009a}]{johnson09}
{Johnson} R.,  {Ponman} T.~J.,    {Finoguenov} A.,  2009a, \mnras, 395, 1287

\bibitem[\protect\citeauthoryear{{Johnson}, {Ponman}, {Rasmussen} \&
  {Finoguenov}}{{Johnson} et~al.}{2009b}]{johnson09b}
{Johnson} R.,  {Ponman} T.~J.,  {Rasmussen} J.,    {Finoguenov} A.,  2009b,
  submitted to \mnras

\bibitem[\protect\citeauthoryear{{Kaiser} \& {Binney}}{{Kaiser} \&
  {Binney}}{2003}]{kaiser03}
{Kaiser} C.~R.,  {Binney} J.,  2003, \mnras, 338, 837

\bibitem[\protect\citeauthoryear{{Leccardi} \& {Molendi}}{{Leccardi} \&
  {Molendi}}{2008}]{leccardi08}
{Leccardi} A.,  {Molendi} S.,  2008, \aap, 487, 461

\bibitem[\protect\citeauthoryear{{McCarthy}, {Babul}, {Bower} \&
  {Balogh}}{{McCarthy} et~al.}{2008}]{mccarthy08}
{McCarthy} I.~G.,  {Babul} A.,  {Bower} R.~G.,    {Balogh} M.~L.,  2008,
  \mnras, 386, 1309

\bibitem[\protect\citeauthoryear{{McNamara}, {Kazemzadeh}, {Rafferty},
  {B{\^\i}rzan}, {Nulsen}, {Kirkpatrick} \& {Wise}}{{McNamara}
  et~al.}{2009}]{mcnamara09}
{McNamara} B.~R.,  {Kazemzadeh} F.,  {Rafferty} D.~A.,  {B{\^\i}rzan} L.,
  {Nulsen} P.~E.~J.,  {Kirkpatrick} C.~C.,    {Wise} M.~W.,  2009, \apj, 698,
  594

\bibitem[\protect\citeauthoryear{{McNamara}, {Nulsen}, {Wise}, {Rafferty},
  {Carilli}, {Sarazin} \& {Blanton}}{{McNamara} et~al.}{2005}]{mcnamara05}
{McNamara} B.~R.,  {Nulsen} P.~E.~J.,  {Wise} M.~W.,  {Rafferty} D.~A.,
  {Carilli} C.,  {Sarazin} C.~L.,    {Blanton} E.~L.,  2005, \nat, 433, 45

\bibitem[\protect\citeauthoryear{{Nipoti} \& {Binney}}{{Nipoti} \&
  {Binney}}{2005}]{nipoti05}
{Nipoti} C.,  {Binney} J.,  2005, \mnras, 361, 428

\bibitem[\protect\citeauthoryear{{Nulsen}, {McNamara}, {Wise} \&
  {David}}{{Nulsen} et~al.}{2005}]{nulsen05}
{Nulsen} P.~E.~J.,  {McNamara} B.~R.,  {Wise} M.~W.,    {David} L.~P.,  2005,
  \apj, 628, 629

\bibitem[\protect\citeauthoryear{{Oh} \& {Benson}}{{Oh} \&
  {Benson}}{2003}]{oh_benson}
{Oh} S.~P.,  {Benson} A.~J.,  2003, \mnras, 342, 664

\bibitem[\protect\citeauthoryear{{O'Neill}, {DeYoung} \& {Jones}}{{O'Neill}
  et~al.}{2009}]{oneill09}
{O'Neill} S.~M.,  {DeYoung} D.~S.,    {Jones} T.~W.,  2009, \apj, 694, 1317

\bibitem[\protect\citeauthoryear{{Parrish} \& {Quataert}}{{Parrish} \&
  {Quataert}}{2008}]{parrish08a}
{Parrish} I.~J.,  {Quataert} E.,  2008, \apjl, 677, L9

\bibitem[\protect\citeauthoryear{{Parrish}, {Quataert} \& {Sharma}}{{Parrish}
  et~al.}{2009}]{parrish09}
{Parrish} I.~J.,  {Quataert} E.,    {Sharma} P.,  2009, preprint
  (arXiv:0905.4500)

\bibitem[\protect\citeauthoryear{{Parrish} \& {Stone}}{{Parrish} \&
  {Stone}}{2005}]{parrish05}
{Parrish} I.~J.,  {Stone} J.~M.,  2005, \apj, 633, 334

\bibitem[\protect\citeauthoryear{{Parrish} \& {Stone}}{{Parrish} \&
  {Stone}}{2007}]{parrish07}
{Parrish} I.~J.,  {Stone} J.~M.,  2007, \apj, 664, 135

\bibitem[\protect\citeauthoryear{{Parrish}, {Stone} \& {Lemaster}}{{Parrish}
  et~al.}{2008}]{parrish08}
{Parrish} I.~J.,  {Stone} J.~M.,    {Lemaster} N.,  2008, \apj, 688, 905

\bibitem[\protect\citeauthoryear{{Peterson} \& {Fabian}}{{Peterson} \&
  {Fabian}}{2006}]{peterson_fabian}
{Peterson} J.~R.,  {Fabian} A.~C.,  2006, \physrep, 427, 1

\bibitem[\protect\citeauthoryear{{Poole}, {Babul}, {McCarthy}, {Sanderson} \&
  {Fardal}}{{Poole} et~al.}{2008}]{poole08}
{Poole} G.~B.,  {Babul} A.,  {McCarthy} I.~G.,  {Sanderson} A.~J.~R.,
  {Fardal} M.~A.,  2008, \mnras, 391, 1163

\bibitem[\protect\citeauthoryear{{Quataert}}{{Quataert}}{2008}]{quataert08}
{Quataert} E.,  2008, \apj, 673, 758

\bibitem[\protect\citeauthoryear{{Rasmussen} \& {Ponman}}{{Rasmussen} \&
  {Ponman}}{2007}]{rasmussen07}
{Rasmussen} J.,  {Ponman} T.~J.,  2007, \mnras, 380, 1554

\bibitem[\protect\citeauthoryear{{Rebusco}, {Churazov}, {B{\"o}hringer} \&
  {Forman}}{{Rebusco} et~al.}{2005}]{rebusco05}
{Rebusco} P.,  {Churazov} E.,  {B{\"o}hringer} H.,    {Forman} W.,  2005,
  \mnras, 359, 1041

\bibitem[\protect\citeauthoryear{{Ruszkowski} \& {Begelman}}{{Ruszkowski} \&
  {Begelman}}{2002}]{ruszkowski02}
{Ruszkowski} M.,  {Begelman} M.~C.,  2002, \apj, 581, 223

\bibitem[\protect\citeauthoryear{{Ruszkowski}, {En{\ss}lin}, {Br{\"u}ggen},
  {Heinz} \& {Pfrommer}}{{Ruszkowski} et~al.}{2007}]{ruszkowski07}
{Ruszkowski} M.,  {En{\ss}lin} T.~A.,  {Br{\"u}ggen} M.,  {Heinz} S.,
  {Pfrommer} C.,  2007, \mnras, 378, 662

\bibitem[\protect\citeauthoryear{{Sanderson}, {O'Sullivan} \&
  {Ponman}}{{Sanderson} et~al.}{2009}]{sanderson09}
{Sanderson} A.~J.~R.,  {O'Sullivan} E.,    {Ponman} T.~J.,  2009, \mnras, 395,
  764

\bibitem[\protect\citeauthoryear{{Sanderson}, {Ponman} \&
  {O'Sullivan}}{{Sanderson} et~al.}{2006}]{sanderson06}
{Sanderson} A.~J.~R.,  {Ponman} T.~J.,    {O'Sullivan} E.,  2006, \mnras, 372,
  1496

\bibitem[\protect\citeauthoryear{{Scannapieco} \& {Oh}}{{Scannapieco} \&
  {Oh}}{2004}]{evan_oh}
{Scannapieco} E.,  {Oh} S.~P.,  2004, \apj, 608, 62

\bibitem[\protect\citeauthoryear{{Sharma}, {Chandran}, {Quataert} \&
  {Parrish}}{{Sharma} et~al.}{2009}]{sharma09}
{Sharma} P.,  {Chandran} B.~D.~G.,  {Quataert} E.,    {Parrish} I.~J.,  2009,
  \apj, 699, 348

\bibitem[\protect\citeauthoryear{{Stone} \& {Norman}}{{Stone} \&
  {Norman}}{1992}]{stone92}
{Stone} J.~M.,  {Norman} M.~L.,  1992, \apjs, 80, 753

\bibitem[\protect\citeauthoryear{{Voit} \& {Donahue}}{{Voit} \&
  {Donahue}}{2005}]{voit05}
{Voit} G.~M.,  {Donahue} M.,  2005, \apj, 634, 955

\bibitem[\protect\citeauthoryear{{Zakamska} \& {Narayan}}{{Zakamska} \&
  {Narayan}}{2003}]{zakamska03}
{Zakamska} N.~L.,  {Narayan} R.,  2003, \apj, 582, 162

\end{thebibliography}
\label{lastpage}

\end{document}